\documentclass[12pt]{article}
\usepackage{graphicx}

\newsavebox{\sboxpubnumber}
\newsavebox{\sboxpubdate}

\newcommand{\Title}[1]{\begin{center} {\Large #1 } \end{center}}
\newcommand{\Author}[1]{\begin{center}{ \sc #1} \end{center}}
\newcommand{\Address}[1]{\begin{center}{ \it #1} \end{center}}

\newenvironment{Abstract}{\begin{quotation}  }{\end{quotation}}
\newenvironment{Presented}{\begin{quotation} \begin{center}
             PRESENTED AT\end{center}\bigskip
      \begin{center}\begin{large}}{\end{large}\end{center}
      \end{quotation}}
\newcommand{\Acknowledgements}{\bigskip  \bigskip \begin{center} \begin{large}
             \bf ACKNOWLEDGEMENTS \end{large}\end{center}}

\newcommand{\la}{\mathrel{\hbox{\rlap{\hbox{\lower4pt
\hbox{$\sim$}}}\hbox{$<$}}}}

\newcommand{\ga}{\mathrel{\hbox{\rlap{\hbox{\lower4pt
\hbox{$\sim$}}}\hbox{$>$}}}}

\begin{document}

\begin{titlepage}

\vfill
\Title{The influence of free neutrons on dynamics and radiation
of astrophysical plasmas}
\vfill
\Author{A.\,A.\,Belyanin, E.\,V.\,Derishev,\newline
V.\,V.\,Kocharovsky, and Vl.\,V.\,Kocharovsky}
\Address{Institute of Applied Physics
of the Russian Academy of Sciences \\ 46 Ulyanov St., 603950 Nizhny Novgorod,
Russia} \vfill
\begin{Abstract}
We present arguments in favor of the presence of free neutrons in plasmas generated by compact astrophysical
objects and find conditions necessary for the formation of the neutron component. The broad range of phenomena
caused by neutrons includes both dynamical (sources' variability, transition of fireballs to the two-flow regime)
and kinetic (fission of helium nuclei by neutrons, electromagnetic cascade, emission in annihilation and nuclear
lines, neutrino losses) effects. The presented theory can be applied to internal regions of accretion disks, jets
in microquasars, and gamma-ray burst (GRB) fireballs.
\end{Abstract}
\vfill
\begin{Presented}
    COSMO-01 \\
    Rovaniemi, Finland, \\
    August 29 -- September 4, 2001
\end{Presented}
\vfill
Also published in:\qquad {\em Radiophys. and Quantum Electron.},
 {\bf 44}, \# 1-2, 3 (2001).
\end{titlepage}
\def\thefootnote{\fnsymbol{footnote}}
\setcounter{footnote}{0}

\section{Introduction}

The lifetime of a free neutron is small by astrophysical standards and amounts to only
about 900~s. Therefore, neutrons can affect significantly the plasma dynamics and
kinetics only in such objects in which the decay of free neutrons is compensated by
processes of neutron accumulation. In fact, there are two such processes: fission of
helium (or nuclei of heavier elements) and proton-to-neutron conversion due to the weak
or, at very high energies, electromagnetic interaction. Of course, we do not touch on the
specific case of superdense matter inside neutron stars where neutrons are practically
stable and their role is dominant in many respects.

Very little attention was paid up to now to the role of free neutrons in the dynamics and
radiation of nondegenerate astrophysical plasmas. This can be explained in part by the
widely accepted view that unstable particles cannot influence significantly the majority
of processes in a plasma except fast nuclear reactions. In this respect, we should
especially emphasize the synthesis of primordial helium in the early Universe, in which
free neutrons play the key role. However, this has little in common with problems of
plasma physics itself. The same is true as far as the nucleosynthesis inside ordinary
stars is concerned.

The few exceptions from this tradition are the following. A number of authors
(e.g.,~\cite{CK}) considered neutrons created due to inelastic interaction of photons and
protons ($p + \gamma \rightarrow n+ \pi^{+}$) as a species, independent of the
magnetic-field influence, which is capable of transferring energy in active galactic
nuclei. In this case, the typical energies of particles exceed $10^{15}$~eV, i.e.,
problems of this kind are related to the physics of cosmic rays and thus suffer
ambiguities in formulation and models adopted.

In~[2, 3], advection-dominated accretion flows (ADAFs) in which ion temperature is very high were proposed as
another class of sources where free neutrons could appear. In this case, neutrons are created due to fission of
helium nuclei in a hot nonisothermal plasma~[4, 5]. In the cited papers, the role of neutrons was only discussed
in relation to generation of gamma-ray lines after neutron capture by heavy elements or generation of
bremsstrahlung radiation due to proton-neutron collisions~\cite{AS}, though the effect of neutron component on the
plasma viscosity was considered, as well~[6].

In this paper, we consider a wider class of phenomena related to the existence of free
neutrons in plasmas created by compact objects. It will be shown that the list of objects
in which neutrons appear in a natural way should be extended to include also relativistic
jets in microquasars~[7] and gamma-ray bursts. On the other hand, the conditions of the
formation of the neutron component appear to be much less stringent than was believed.

Let us recall the main properties of gamma-ray bursts (see also the reviews~[8, 9]). The
burst sources are distributed isotropically in the celestial sphere, have angular sizes
less than the angular resolution of modern detectors, and produce (probably nonrecurrent)
bursts of durations $0.1\!-\!100$~s. The maximum of the spectrum falls in the gamma-ray
band (about 200~keV), which explains the name of this phenomenon. Recent observations of
optical components of gamma-ray bursts~[10--12] revealed that GRB sources are located at
distances comparable to the size of the Universe. The total energy release of a typical
source amounts to $10^{51}\!-\!10^{52}$~ergs. Consequently, the observed radiation should
be generated in an ultrarelativistic flow of matter moving with a Lorentz factor $\Gamma
>100$~\cite{gamma}, which allows one to avoid strong two-photon absorption.

When discussing the physics of a two-component relativistic flow (fireball), we will
focus on issues not considered in detail in our previous papers~[14, 15]. In particular,
we mean the secondary dissociation of helium, the influence of the magnetic field on
neutrino losses, and the spectrum of suprathermal emission of a photosphere in a wide
energy range.

For simplicity, we assume that the fireball is isotropic. However, all the results presented remain valid
also for jets with opening angles and Lorentz factors satisfying the condition $\theta > 1/\Gamma$. In the case of
a jet, the source luminosity, mass-loss rate, etc., mean the corresponding values per unit solid angle multiplied
by $4\pi$.

\section{Neutrons in accretion disks}

Fission of helium nuclei due to inelastic collisions with high-energy ions in an
accretion disk or with neutrons penetrating it is the main source of neutrons in
accretion disks. It is usually assumed that the created neutrons are advected by the bulk
plasma motion toward the central compact object\,\footnote {That is, toward a neutron
star or a black hole. Hereafter, for definiteness, we will mean a black hole.} with a
velocity equal to the radial disk velocity~[4, 5]. However, if the accretion-disk plasma
is tenuous enough, dynamical decoupling of the neutron and ion components of the disk
takes place, which results in accumulation of neutrons. The term ``decoupling'' means the
appearance of a non-negligible difference in the hydrodynamical velocities of the two
components.

\subsection{Pile-up of free neutrons in a disk}

Consider the radial hydrodynamical motion of neutrons caused by the angular momentum loss
due to neutron-ion (mostly, neutron-proton) collisions inside an accretion disk.
Straightforward calculation yields
\begin{equation}
V_{\rm n}^{(r)}=- \frac{\rm d}{{\rm d}t} \left( \frac{R_{\rm g}c^2}{2V_{\rm n}^2}
\right)= \frac{R_{\rm g}c^2}{V_{\rm n}^3} \frac{{\rm d} V_{\rm n}}{{\rm d}t} =
- \frac{R_{\rm g}c^2}{2V_{\rm n}^3} \nu_{\rm pn} \left( V_{\rm n} -
V_{\rm d}\right),
\end{equation}
where $V_{\rm n}$ and $V_{\rm n}^{(r)}$ are the orbital and radial velocities of the neutron component, $R_{\rm
g}$ is the Schwarzschild radius of the black hole, and $c$ is the speed of light. In the last expression, the time
derivative of $V_{\rm n}$ was replaced by $- \nu_{\rm pn}\,(V_{\rm n} - V_{\rm d})/2$, where $\nu_{\rm pn}$ is the
rate of neutron-proton collisions and $V_{\rm d}$ is the orbital velocity of the disk.

The orbital velocity of the neutron component exceeds that of the disk, so that the excess centrifugal force
should be balanced by the friction force due to the radial-velocity difference:

\begin{equation}
\frac{V_{\rm n}^2 - V_{\rm d}^2}{R} =
 \frac{\nu_{\rm pn}}{2} \left( V_{\rm n}^{(r)} - V_{\rm d}^{(r)} \right).
\end{equation}
Here $V_{\rm d}^{(r)}$ is the radial velocity of the disk and $R$ is the current radius. Assuming $V_{\rm n}
\approx V_{\rm d} \approx V_0$, where $V_0$ is the Keplerian velocity, we obtain:
\begin{equation}
V_{\rm n}^{(r)} = - \frac{1}{4} \left( \frac{\nu_{\rm pn}R}{V_0}\right)^2
\left( V_{\rm n}^{(r)} - V_{\rm d}^{(r)} \right).
\end{equation}
The quantity $\nu_{\rm pn}R/V_0 $ is equal to the number of collisions of a neutron during a single orbital
rotation divided by $2\pi$.

The collision rate $\nu_{\rm pn}$ can be expressed in terms of the optical depth for
neutrons $\tau_{\rm d}$, which is equal to the average number of collisions of a neutron
penetrating the accretion disk perpendicular to its plane with a velocity about the
Keplerian velocity. Assuming that the thickness of the ion and neutron ``disks'' are
$h_{\rm d}$ and $h_{\rm n}$, respectively, and that the cross-section of neutron-proton
collisions is inversely proportional to their relative velocities, we find $\nu_{\rm pn}
= \langle \sigma V \rangle_{\rm pn}\,n_{\rm p}\,(h_{\rm d}/h_{\rm n})=(V_0/h_{\rm
n})\,(\left< \sigma V \right>_{\rm pn} /V_0)\,n_{\rm p}h_{\rm d} =\tau_{\rm
d}V_0/(2h_{\rm n})$. Here $n_{\rm p}$ and $n_{\rm n}$ are the number densities of protons
and neutrons, respectively, and $\langle\sigma V\rangle_{\rm pn}$ is the product of the
proton-neutron cross section and their relative velocity averaged over the distribution
function. Finally, the expression for the radial velocity of neutrons takes the following
form :
\begin{equation}
V_{\rm n}^{(r)} = \frac{[R/(2h_{\rm n})]^2\,\tau_{\rm d}^2}{4+[R/(2h_{\rm n})]^2\,
\tau_{\rm d}^2}V_{\rm d}^{(r)}.
\end{equation}
The neutron pile-up and the formation of a neutron halo in the inner part of an accretion disk begins if $V_{\rm
n}^{(r)}$ becomes much less than $V_{\rm d}^{(r)}$, i.e., $\tau_{\rm d} \la 4 h_{\rm n}/R$. Correspondingly,
the accretion rate should be less than
\begin{equation}
\dot{M}_{\max} =
\frac{8 \pi V_{\rm d}^{(r)} V_0 h_{\rm n} m_{\rm p}}
{\langle \sigma V \rangle_{\rm pn}}=
\frac{4 \pi \alpha R_{\rm g} c^2}{\langle \sigma V \rangle_{\rm pn}}
\left( \frac{h_{\rm d}}{R} \right)^{2} \left(\frac{h_{\rm n}}{R}\right) m_{\rm
p},
\end{equation}
where $m_{\rm p}$ is the proton mass. Here we used the relation $V_{\rm
d}^{(r)}=\alpha\,(h_{\rm d}/R)^2\,V_0$ commonly accepted for $\alpha$-disk models, while
$h_{\rm d}/R$ plays the role of the model parameter. If $\alpha =0.1$ and $h_{\rm d}
\approx R$, the limiting accretion rate $\dot{M}_{\max}$ corresponds to a luminosity
about the Eddington one.

As the accretion rate decreases, the role of the effect of neutron pile-up becomes more important, and eventually
the number density of neutrons exceeds that of protons and the proper viscosity of the neutron component becomes
important. Depending on the adopted disk model, a further decrease in the accretion rate results in either
stabilization of the ratio $V_{\rm n}^{(r)}/V_{\rm d}^{(r)}$ or the sign reversal of $V_{\rm n}^{(r)}$ and
subsequent tending of its absolute value to zero.

\subsection{Virialization of neutrons in the halo}

A neutron orbit is subject to precession caused by the gravimagnetic force by the rotating black hole. The
precession is an effective mechanism of additional heating of neutrons and can result in the complete
virialization of the neutron halo, which is possible if the angular velocity of the precession exceeds the rate of
neutron-proton collisions, i.e.,
\begin{equation}
\frac{ac}{2\pi R_{\rm g}} \left( \frac{R_{\rm g}}{R} \right)^3
\ga \frac{\nu_{\rm pn}}{2}\,,
\end{equation}
where $a$ is the angular momentum of the black hole in units of $M R_{\rm g} c/2$.
Substituting the expression for $\nu_{\rm pn}$ in which the proton number density $n_{\rm
p}$ is given in terms of $\dot{M}$ and $R$, one can easily obtain the criterion for the
accretion rate:

\begin{equation}\label{virial}
\dot{M} \la \frac{a}{\sqrt{2}\,\pi}
\left( \frac{R_{\rm g}}{R} \right)^{3/2} \dot{M}_{\max}\,.
\end{equation}
It is clear that the considered mechanism can hardly provide virialization near the outer
edge of the neutron halo even in the case of a rapidly rotating black hole ($a \to 1$).
Nevertheless, virialization is possible in the inner part of the halo at
distance\begin{equation} \label{virial2} R
\la\left(\frac{a\dot{M}_{\max}}{\sqrt{2}\,\pi\dot{M}}\right)^{2/3} R_{\rm g}\,.
\end{equation}

\subsection{Long-term variability}

Neutron-halo virialization near the black hole results in the formation of an almost
isotropic wind that carries away about one-half of the neutrons created due to helium
fission in the accretion disk and moves at a velocity comparable to the speed of light.
Neutrons in the wind decay, and the resulting protons exert dynamical pressure on the
accreted matter. At a distance less than $t_{\rm n} V_{\rm n}$, where $V_{\rm n}$ is the
wind velocity and $t_{\rm n}$ is the lifetime of a neutron, this pressure is equal to
\begin{equation}
P_w \approx \frac{\eta_{\rm He}\dot{M}}{16\pi R t_{\rm n}}\,,
\end{equation}
where $\eta_{\rm He}$ is the mass abundance of helium. In the case of quasi-spherical
accretion (e.g., from a stellar wind from a massive companion star), the neutron wind can
terminate the accretion if $P_w$ exceeds the dynamical pressure of the infalling matter
at the sonic point $R_{\rm c}\approx (c/V_{\infty})^2\,R_{\rm g}$, where $V_{\infty}$ is
the velocity of the stellar wind at a large distance from the black hole. The
corresponding condition can be presented in the following form:
\begin{equation}
\label{pressure}
\frac{\dot{M} V_{\infty}}{4\pi R_{\rm c}^2}\la P_w
\quad \Rightarrow \quad
\frac{V_{\infty}}{c} \la
\left( \frac{\eta_{\rm He}R_{\rm g}}{4c t_{\rm n}} \right)^{1/3}.
\end{equation}
As soon as the condition (\ref{pressure}) is satisfied, the accretion terminates, but the
source remains active due to the matter accumulated in the disk. This matter will be
exhausted during a time interval of the order of the characteristic time scale of
angular-momentum transfer at the outer edge of the disk. Then the source switches off,
and the next cycle of the accretion from the stellar wind begins. The quiescent period
lasts until the matter from the outer edge of the disk reaches the region of neutron-halo
formation, i.e., it is approximately equal to the active period. The above-described
sequence of events results in the long-term variability of a period very sensitive to the
size of the outer edge of the disk, i.e., to the specific angular momentum of the
accreted matter. In some cases, this period can amount to a few hundred or thousand
years.

\subsection{Hard gamma-ray emission from the neutron halo}

If the neutron halo is virialized, pions can be created at distance $R \sim 3 R_{\rm g}$ due to neutron-proton and
neutron-neutron collisions. Assuming that the average rate of pion creation $\xi_{\pi} = \langle \sigma_{\pi}
V\rangle$ averaged over velocity distributions of the nucleons is the same for both types of collisions, we find
the total ``pion luminosity''
\begin{equation}
\label{pi1}
L_{\pi} \sim \xi_{\pi}\left(\frac{n_{\rm n}^2}{2}+
\frac{h_{\rm d}}{h_{\rm n}} n_{\rm n} n_{\rm p}\right) R_{\pi}^3 E_{\pi},
\end{equation}
where $R_{\pi}$ is the distance at which pion creation becomes possible and $E_{\pi} \approx 140$~MeV is the pion
rest energy. If helium nuclei dissociate completely, the neutron number density can be determined using the
balance condition\begin{equation} \label{pi2} \frac{\eta_{\rm He} \dot{M}}{2 m_{\rm p}} = \xi_{\rm loss}
\frac{n_{\rm n}^2}{2} R_{\pi}^3,
\end{equation}
where the loss rate of neutron $\xi_{\rm loss}$ is determined mainly by their scattering into the black hole.
Determining $n_{\rm n}$ from Eq.~(\ref{pi2}) and $n_{\rm p}$ from the continuity equation and assuming $h_{\rm n}
= R$, we obtain the ``pion luminosities'' of neutron-neutron and proton-neutron collisions:
\begin{equation} \label{pi3}
L_{\rm nn} = \frac{\xi_{\pi}}{\xi_{\rm loss}} \frac{\eta_{\rm He} \dot{M} E_{\pi}}{2 m_{\rm p}}\,,
\end{equation}
\begin{equation} \label{pi4}
L_{\rm pn} = \frac{E_{\pi}}{2\pi \alpha} \frac{\xi_{\pi}}{\xi_{\rm loss}}
\left( \frac{R}{h_{\rm d}}\right)^2 \left( \frac{\eta_{\rm He} \xi_{\rm
loss} \dot{M}^3}{R_{\rm g} c^2  m_{\rm p}^3}\right)^{1/2}.
\end{equation}
The limiting values of these luminosities
\begin{equation} \label{pi5}
L_{\rm nn}^{\max} \approx 7\cdot 10^{38} \alpha \eta_{\rm He} M_{10}
 \left(\frac{h_{\rm d}}{R}\right)^2~{\rm erg/s},
\end{equation}
\begin{equation} \label{pi6}
L_{\rm pn}^{\max} \approx 10^{38} M_{10}\,\sqrt{\alpha \eta_{\rm He}} \left(\frac{h_{\rm
d}}{R}\right)~{\rm erg/s}
\end{equation}  can be obtained assuming $\dot{M} =
\dot{M}_{\max}$. Here $M_{10}$ is the mass of the black hole in units of $10M_{\odot}$,
where $M_{\odot}$ is the solar mass. About one-half of the pion energy is transformed to
$\gamma$-quanta and electron-positron $(e^{-}e^{+})$ pairs. This energy fraction is
reprocessed by virtue of comptonization of soft X-ray radiation and the electromagnetic
cascade.

Electrons and positrons cool rapidly due to interaction with the X-ray radiation of the
disk, and the equilibrium distribution of $e^{-}e^{+}$ pairs is characterized by a mean
energy of about $50\!-\!500$~keV. The spectrum of photons with energies above a few tens
of keV (i.e., above the cutoff of the blackbody emission from the disk) has a typical
comptonization profile. The electromagnetic cascade, if it develops, leads to the
formation of a hard power-law ``tail'' with an index of about 2 (we mean the photon
number per unit frequency range). Both spectral features are frequently observed during
hard states of accreting sources --- black hole candidates. Moreover, one should expect
hard radiation with photon energies up to 70~MeV which, however, is hardly detectable
since it carries much less energy compared to the hard X-ray radiation.

The neutron halo can also be an efficient source of radiation in the deuterium line at energy $E_{\rm d} \approx
2.2$~MeV. Deuterium is created due to evaporation of neutrons from the halo and their subsequent radiative capture
by protons in the cold outer disk. The total luminosity in the line is equal to $\beta \eta_{\rm He} \dot{M}
E_{\rm d}/(2 m_{\rm p})$, where $\beta$ is the fraction of neutrons entering the outer disk. This fraction is
strongly dependent on the velocity distribution of the neutrons and the halo size.

\section{Neutrons in relativistic flows}

In this section, we discuss the characteristics of the neutron component in relativistic
(and super-Eddington) plasma flows from compact objects. We call an object compact if the
characteristic dynamical time, which is approximately equal to $3 R_{\rm g} /c$, is small
compared to the neutron lifetime. This condition excludes from our analysis supermassive
black holes in active galactic nuclei if their mass exceeds $3 \cdot 10^7 M_{\odot}$.
Moreover, for the sources with moderate densities and temperatures at the wind base,
e.g., microquasars, some seed neutrons are also necessary. A neutron halo can be a source
of such seed neutrons.

\subsection{Conditions for the formation and existence of the neutron component}

Gamma-ray bursts are probably the only example of sources in which the physical
conditions necessary for independent (spontaneous) appearance of neutrons in a
relativistic wind are met. For a nucleon density in the wind base
$10^4\!-\!10^6$~g/cm$^3$ typical for gamma-ray bursts, the dissociation temperature is
about 0.7~MeV for helium and is even less for other nuclei. In fact, the temperature
$T_0$ in the source is about an order of magnitude higher and amounts to $3\!-\!10$~MeV.
According to the generally accepted point of view, ultrarelativistic plasma flows giving
rise to gamma-ray bursts are formed during either catastrophic events in neutron-star
evolution or the collapse of central parts of supermassive stars~[16--20]. Therefore, the
matter in the base of the fireball is initially enriched by neutrons, and even ordinary
dissociation of the nuclei composing this matter makes the ratio of number densities of
neutrons and protons amount to $n_{\rm n}/n_{\rm p}\ga 1$.

There is another mechanism leading to creation of free neutrons in gamma-ray burst
sources, namely, mutual conversion of protons and neutrons due to the following
weak-interaction processes: $p+\bar{\nu}_{\rm e} \to n + e^{+}$ and $p+e^{-} \to n +
\nu_{\rm e}$. A time of about~[21]
\begin{equation} \label{eqv}
t_w \sim \frac{60\pi^3 \hbar^7 c^6}{Q^5 G_{\rm F}^2\,(1+3g_{\rm A}^2)}\, \approx 3\cdot
10^{-2}\left(\frac{Q}{10~{\rm MeV}}\right)^{-5}~{\rm s}
\end{equation} is enough for
thermal equilibrium to be established, in which the number densities of protons and
neutrons are approximately equal. Here $G_{\rm F} \approx 10^{-49}$~erg\,$\cdot$\,cm$^3$
is the Fermi constant, $Q$ is the typical energy of interacting particles ($Q\approx 3\,
T_0$ for a thermal distribution function), and $g_{\rm A} \approx 1.25$. In the course of
the wind acceleration, the temperature decreases inversely proportional to the radius,
and the time scale for establishing equilibrium increases sharply. Quite soon, at a
temperature exceeding $(m_{\rm n} -m_{\rm p}) c^2$, the density ratio becomes independent
of the temperature behavior and remains equal to $\eta \equiv n_{\rm n}/n_{\rm
p}\approx 1$.

When the plasma temperature falls to about 70~keV, the recombination of protons and
neutrons becomes possible. The process comprises two stages. First, a proton and a
neutron form a deuteron via the reaction $p+n \rightarrow d+\gamma$ and then the reaction
chain splits in different channels leading to the synthesis of helium according to the
resulting reaction $3d \rightarrow ^4\!\!\!{\rm He}+p+n$. The recombination rate is
determined mainly by the rate of deuterium synthesis. The recombination is negligible
under the condition $\tau_{\rm r}=2\,\langle \sigma v \rangle_{\rm d}\,n_{\rm p}/3 <
c/(R_{\rm r}\Gamma_{\rm r})$, where $\langle\sigma v \rangle_{\rm d} \approx 5\cdot
10^{-20}$~cm$^3$/s is the constant of the deuterium-synthesis rate, $R_{\rm r}$ is the
distance from the source at which recombination becomes possible, and $\Gamma_{\rm r}$ is
the Lorentz factor at this distance.

The Lorentz factor of the fireball increases almost linearly, $\Gamma \approx R/R_0$,
until its maximum $\Gamma_{\ell}=L/(\dot{M} c^2)$ determined by the source luminosity $L$
and the rate of baryonic mass loss $\dot{M}$. The radius $R_0$ in the fireball base is
assumed to be about $3R_{\rm g}$. To calculate the recombination rate of protons and
neutrons, we use the continuity equation that determines the dependence of the proton
number density $n_{\rm p}$ on the radius:
\begin{equation}
n_{\rm p}=\frac{L}{4\pi R^2\,(1+\eta)\,\Gamma \Gamma_{\ell} m c^3}\,,
\end{equation}
where $m$ is the nucleon mass. Since a temperature of 70~keV is reached in gamma-ray burst fireballs at the
acceleration stage, it can be easily shown that the recombination becomes significant if the following condition
is met:
\begin{equation} \label{rec-cond}
\tau_{\rm r}\approx\left(\frac{R_0}{1~{\rm km}}\right)\left(\frac{T_0}{1.5~{\rm MeV}}\right)
\frac{1}{(1+\eta)\,\Gamma_{\ell}}>1.
\end{equation}
If $\eta \approx 1$, then the fraction of free neutrons left after the recombination is $(\tau_{\rm r} +1)^{-1}$
as long as $\tau_{\rm r}+1<\eta/|1-\eta|$, while at larger values of $\tau_{\rm r}$ this fraction becomes
exponentially small (if $\eta < 1$) or constant (if $\eta > 1$).

\subsection{Decoupling of proton and neutron components}

When the neutron component in a fireball is considered, we focus on the case where
neutrons at the base of the fireball collide with protons frequently enough to be
advected by the bulk flow. In fact, this means that the time between collisions of
neutrons with other nucleons is less than the acceleration time scale $R/(c\Gamma)$,
i.e.,
\begin{equation} \label{drag}
\sigma_{\rm np} v n_{\rm p} R > c\Gamma \quad \Rightarrow \quad
L \ga 12\,(1+\eta)\,\Gamma_{\ell}\frac{\sigma_{\rm T}}{\sigma_{\rm np}}L_{\rm edd}.
\end{equation}
The expression on the right-hand side is calculated at the base of the fireball where
$\Gamma \approx 1$ and $R \approx 3 R_{\rm g}$. Also, we take into account that the
relative velocity of protons and neutrons and the cross-section of their collisions at
the threshold of decoupling are given by $v \sim c/2$ and $\sigma_{\rm np} \approx 6
\cdot 10^{-26}$~cm$^2$, respectively. In the obtained formula, the luminosity is related
to the Eddington limit $L_{\rm edd} = 2\pi m c^3 R_{\rm g}/ \sigma_{\rm T} \approx 1.3
\cdot 10^{38}(M/M_{\odot})$~erg/s, where $\sigma_{\rm T} \approx 6.6 \cdot
10^{-25}$~cm$^2$ is the Thomson cross-section. Note that hereafter the quantity $L$ may
not necessarily be equal to the actual luminosity of the object and may significantly
exceed the latter quantity.

It is clear that the neutrons are indeed advected by the plasma outflow and form a
neutral component if the luminosity $L$ is much greater than the Eddington luminosity.
However, in the course of acceleration and expansion of the wind, the decrease in the
neutron-proton collision rate\,\footnote{To unify terminology, we write ``proton''
though, bearing in mind that helium is created due to the recombination, the term ``ion''
should be used in some cases hereafter.} is much more rapid than the increase in the
acceleration timescale. As a result, upon reaching certain distance, neutrons move almost
freely. If the fireball acceleration continues beyond this distance, the Lorentz factor
of the proton component exceeds that of the neutron component, i.e., decoupling takes
place. The criterion of the decoupling is identical to Eq.~(\ref{drag}) calculated at the
distance $R_{\rm s} \sim \Gamma_{\ell} R_0$ from the source, where the fireball
acceleration is terminated within the framework of the one-fluid approximation. The
obtained condition can be presented in a more illustrative form:
\begin{equation} \label{decoup}
\Gamma_{\ell} \ga \Gamma_* = T_0/ T_*, \qquad T_* \approx 5~{\rm keV},
\end{equation}
where $T_0$ is the temperature at the base of the fireball.

For a given luminosity, the distance at which the decoupling takes place decreases with decreasing $\dot{M}$.
Therefore, the terminal Lorentz factor $\Gamma_{\rm n}$ of the neutron component as a function of $\Gamma_{\ell}$
reaches the maximum at $\Gamma_{\ell} \approx \Gamma_*$ and then decreases as $\Gamma_{\ell}^{-1/3}$. The maximum
value is approximately equal to $\Gamma_*$. The limiting Lorentz factor $\Gamma_{\rm p}$ of the proton component
increases monotonically with increasing $\Gamma_{\ell}$ and, if neutrino losses are negligible (see below), can be
calculated using the energy conservation law: $\Gamma_{\rm p}=\Gamma_{\ell}+\eta\,(\Gamma_{\ell}-\Gamma_{\rm n})$.
The value of $\Gamma_{\rm n}$ can be approximately calculated using the formula
\begin{equation} \label{constr}
\Gamma_{\rm n} \approx \frac{\Gamma_{\rm p} \Gamma_*^{4/3}} {\left( \Gamma_*^4
+2.37\,\Gamma_{\rm p}^4 \right)^{1/3}}\,.
\end{equation}

In the case of gamma-ray bursts, the value of the decoupling parameter $\Gamma_{\rm
p}/\Gamma_{\rm n}$ ranges typically from 1 to 10, which is much less than the limiting
Lorentz factor of the proton flow. On the contrary, the decoupling parameter is close to
$\Gamma_{\rm p}$ in the case of jets in microquasars, so that the neutrons remain
nonrelativistic.

\subsection{Secondary dissociation of helium}

Consider the case where both the condition of efficient neutron recombination and the
condition of the transition of the fireball to the two-flow regime are satisfied. Then,
prior to the decoupling of the proton and neutron components, secondary dissociation of
helium becomes possible. In the accelerated reference frame comoving with the plasma
flow, the neutron is subject to the inertial force $mc^2\,{\rm d}\Gamma/{\rm d}R$
directed toward the central source. Under the action of this force, the neutrons acquire
a certain regular velocity and the neutron component is heated due to collisions with
protons and helium nuclei.

Let $\nu_{\rm np}$ and $\nu_{{\rm n}\alpha}$ be the rates of elastic collisions of neutrons with protons and
alpha-particles, respectively. Coulomb collisions with electrons cool ions quickly to energies of the order of the
photon temperature $T$, so that ions can be considered motionless compared to neutrons. Therefore, steady-state
values of the regular velocity $v$ and r.m.s.\ velocity $\tilde{v}$ of neutrons satisfy the following equations:
\begin{equation} \label{heat}
\left( \frac{\nu_{\rm np}}{2} + \frac{4\nu_{{\rm n}\alpha}}{5} \right) v =
c^2 \frac{{\rm d}\Gamma}{{\rm d}R}\,, \qquad
\left( \frac{\nu_{\rm np}}{2} + \frac{4\nu_{{\rm n}\alpha}}{25} \right)
\tilde{v}^2 \approx 2 v c^2 \frac{{\rm d}\Gamma}{{\rm d}R}\,.
\end{equation}
Just before a collision, the radial velocity of a typical neutron is $v + \delta v/2$,
where $\delta v=c^2\,({\rm d}\Gamma/{\rm d}R)/(\nu_{\rm np} + \nu_{{\rm n}\alpha})$ is
the velocity increment between collisions. Now it is easy to find the mean energy of
collisions of neutrons with helium nuclei, which is given by
\begin{equation}
\label{coll} E_{{\rm n}\alpha}
=
\frac{2m}{5}\left(\tilde{v}^2+v\,\delta
v+ \frac{(\delta v)^2}{4}\right)\approx
\frac{2}{5}\left(
\frac{2\,(\nu_{\rm np}+\nu_{{\rm n}\alpha})^2} {\left(\displaystyle\frac{\nu_{\rm
np}}{2}+\displaystyle\frac{4\nu_{{\rm n}\alpha}}{5}\right)
\left(\displaystyle\frac{\nu_{\rm np}}{2}+\displaystyle\frac{4\nu_{{\rm
n}\alpha}}{25}\right)}+ \frac{\nu_{\rm np}+\nu_{{\rm
n}\alpha}}{\displaystyle\frac{\nu_{\rm np}}{2}+ \displaystyle\frac{4\nu_{{\rm
n}\alpha}}{5}}+\frac{1}{4}\right) m\,(\delta v)^2\,.
\end{equation}
This value is equal to $4.1\,m\,(\delta v)^2$ and $6.85\,m\,(\delta v)^2$ in the limiting cases of hydrogen and
helium plasma, respectively.

If $E_{{\rm n}\alpha}$ exceeds the threshold of the reaction
$n+{}^4$He$ \rightarrow d+{}^3{\rm H}$, which is about 18~MeV, multiplication
of neutrons due to helium fission becomes possible.\,\footnote{In fact, helium
dissociates completely, ${}^4{\rm He}\rightarrow 2n +2p$, since all other
reactions have lower energy thresholds and greater cross-sections.} In this
case, the total number of neutrons increases by a factor of $3^N$, where $N$
is the number of inelastic collisions per neutron which occur during a time of
about the fireball-expansion time scale $R/(c\Gamma)$. Since the total number
of collisions during this time is about $N_{\rm t}=c/(\delta v)$, we arrive at
the estimate $N=\nu_{\rm diss}N_{\rm t}/ (\nu_{\rm np} +\nu_{{\rm n}\alpha})
\ll N_{\rm t}$.

Consider the case where $\eta >1$ and the recombination proceeded almost completely,
i.e., the fireball at the onset of the secondary dissociation comprises only neutrons and
helium nuclei. In this case, $N_{\rm t}\approx 18$, and the rate of collisions resulting
in the dissociation is equal to $\nu_{\rm diss}\approx 0.1\nu_{{\rm n}\alpha}$, so that
the number of neutrons increases by a factor of 10. Of course, rigorous calculation
should account for the fact that protons are created due to the dissociation, and the
ratio $N/N_{\rm t}$, as well as the value of $N_{\rm t}$, are not constants (both these
quantities decrease). Due to the appearance of protons in the course of the secondary
dissociation, about one-half of the neutrons remains bound in helium nuclei after the
recombination. But if $\eta$ is significantly less than unity but still $\tau_{\rm r} \gg
1$, secondary dissociation is almost impossible.

The effect of secondary dissociation is significant for fireballs in GRB sources. At the
same time, the influence of this effect on the jets in microquasars is negligible since
the plasma in these jets is predominantly hydrogen. We should specially mention the
so-called soft gamma repeaters (SGRs), i.e., neutron stars which produce sporadic bursts
of soft gamma-ray emission of duration about 0.1~s and energy $10^{38}\!-\!10^{43}$~ergs.
No plausible theory of this phenomenon has been proposed up to now. Quite often, however,
SGRs are assumed to be scaled-down analogs of GRBs. In these objects, the temperature is
too low to cause the dissociation of helium at the base of the fireball. At the same
time, low hydrogen abundance should lead to secondary dissociation. Thus, the formation
of the neutron component in SGRs is entirely determined by the rate of nonthermal
processes producing the seed neutrons. One cannot exclude the possibility that free
neutrons play a noticeable role in SGRs, too.

\subsection{Radiative processes related to decoupling}

The collisions of two nucleons may be inelastic if their kinetic energy exceeds the
pion-creation threshold of about 140~MeV. After the decoupling, the proton (ion)
component remains cold, while the neutron one is heated up to subrelativistic
temperatures. Moreover, the relative velocity of protons and neutrons becomes
relativistic, as well. If neutrons dominate the composition of nucleonic matter at the
onset of the decoupling, the number densities of neutrons and protons become roughly
equal, since inelastic collision of two neutrons leads to the conversion of one of these
particles to a proton with a probability close to unity.

As the collision energy increases, the total cross-section of inelastic processes
increases rapidly and becomes comparable to the cross-section of elastic scattering, so
that a nucleon on average undergoes one inelastic collision (one should exclude the case
$\eta \gg 1$ because of the above-mentioned reasons). The probabilities of creation of
neutral or charged pions in proton-neutron collisions are almost equal.\footnote{Among
other inelastic reactions, processes leading to deuterium formation (e.g., $p+n
\rightarrow d + \pi^0$) take place. The created deuterons have a good chance to avoid new
collisions and, therefore, dissociation. As a result, the fireball becomes an efficient
source of deuterium: up to 20\% of the ejected nucleon matter is converted to deuterium.}
The decay of charged pions, $\pi^{+} (\pi^{-}) \rightarrow \mu^{+} (\mu^{-}) + \nu_{\mu}
(\bar{\nu}_{\mu}) \rightarrow e^{+} (e^{-}) + \nu_{\rm e} (\bar{\nu}_{\rm e}) + \nu_{\mu}
+ \bar{\nu}_{\mu}$, results in the appearance of three neutrinos and a positron
(electron), and each particle carries approximately the same energy of about 35~MeV.
Neutral pions decay into a pair of gamma-quanta of energies about 70~MeV.

Therefore, the decoupling leads to neutrino emission, originating mainly in a region near
the saturation radius $R_{\rm s}$, and to the electromagnetic cascade which generates
hard gamma-rays escaping from the photosphere. The distance from the source to the
photosphere, which is located at different radii for photons of different energies, is
generally much greater than the saturation radius, so that the electromagnetic emission
is much weaker than the neutrino one. Nevertheless, it is easier to detect the
electromagnetic signal, while detection of neutrinos is virtually impossible even in the
case of such powerful sources as gamma-ray bursts.

Emission of high-energy particles due to pion decay results in the development of an
electromagnetic cascade in the fireball plasma~\cite{apj}. When analyzing the cascade,
neutral and charged pions can be treated in a unified way. After creation of two
$e^{-}e^{+}$ pairs by photons generated due to the decay of a neutral pion, these species
differ only by the amount of energy transferred to electrons and positrons. Namely, the
photons generated by a neutral-pion decay create two $e^{-}e^{+}$ pairs, and, since the
energies of the daughter particles are roughly equal, they are almost indistinguishable
from the electrons and positrons born in charged-pion decays.

In the case of gamma-ray bursts, the problem is further simplified due to the fact that the cascade develops in
the one-step regime near the photosphere. This means that thermal photons, upon scattering (comptonization) on
initial ultrarelativistic particles, gain enough energy to create secondary pairs that cannot further transfer to
photons the energy necessary for creation of new pairs, and thus the cascade terminates.

What happens to a photon after the comptonization depends on its energy. If this energy
exceeds the cascade threshold $\varepsilon_{\rm t}$, the photon is absorbed by another
photon of smaller energy, and an $e^{-}e^{+}$ pair is created. Otherwise, scattering by
thermal electrons, in which the photon energy decreases gradually, is the dominant
process. The quantity $\varepsilon_{\rm t}$ is universal in the approximation of a
one-step cascade and is about 3~MeV. The newly created electrons and positrons are
decelerated to almost thermal velocities prior to annihilation. As a result, all their
energy is transferred to the background blackbody radiation, which plays the role of a
heat bath with a heat capacity exceeding that of the fireball plasma.

\subsection{Neutrino losses}

Neutrino emission is interesting mostly due to the fact that it can consume a large
fraction of the fireball energy. The neutrino losses increase if the pion momenta are
changed partially or completely before the decay. This is possible, e.g., due to
interaction with the magnetic field. The decay time of a pion, equal to $2.6 \cdot
10^{-8}$~s, exceeds the inverse gyrofrequency if the magnetic field is greater than
600~G. The energy density of such a field is many orders of magnitude smaller than the
thermal energy density of the plasma in the decoupling region. Since the requirements to
the uniformity of the magnetic field are not stringent in this case, we suppose that this
effect is always present. Let us describe the effect of the magnetic field by the
parameter $\theta$, the angle between the field lines and radius. This parameter varies
from 0 to $\pi/2$ for different models of gamma-ray burst sources and jets.

Let us estimate the neutrino losses in the case of large decoupling parameter. If $\Gamma \gg \Gamma_{\rm n}$, all
the three types of pions are created with equal probabilities, and their total energy in the center-of-mass frame
of the colliding nucleons is approximately equal to $mc^2\,\sqrt{\Gamma/\Gamma_{\rm n}}$. Taking into account the
effect of the magnetic field, we find that neutrinos in the laboratory frame carry away the energy
$$\epsilon_{\nu} \approx mc^2 \Gamma^2 \Gamma_{\rm n}^{-1}\displaystyle\frac{\left( 1- \sqrt{1-4\Gamma_{\rm
n}/\Gamma} \cos^2 \theta \right)}{4}\,.$$

The total energy of the emitted neutrinos per proton is equal to
\begin{equation} \label{neutrinos}
E_{\nu} \approx \int\limits_{R_{\rm d}}^{\infty} {\epsilon_{\nu} \sigma_{\rm c} n_{\rm n}
\frac{{\rm d}R}{\Gamma}}\,,
\end{equation}
where $\sigma_{\rm c}\approx 2\,\cdot\,10^{-26}$~cm$^2$ is the cross-section of inelastic
proton-neutron collisions. We divide the integration region into three parts: (i) from
the decoupling radius $R_{\rm d}$ to the point where $\sin \theta = \sqrt{2\Gamma_{\rm
n}/\Gamma}$; (ii) from the previous point to the saturation radius $R_{\rm s}=\Gamma_{\rm
p} R_{\rm d}/\Gamma_{\rm d}$; and (iii) from $R_{\rm s}$ to infinity. With accuracy
sufficient for an estimate, we adopt $n_{\rm n}=\eta n_{\rm p} \Gamma^2/(2\Gamma_{\rm
n}^2)$, where $\Gamma=(R/R_{\rm d})\,\Gamma_{\rm d}$ in regions (i) and (ii), and $\Gamma
=\Gamma_{\rm p}$ in region (iii). If $\sin \theta <\sqrt{2\Gamma_{\rm n}/\Gamma_{\rm
p}}$, then the formula $\epsilon_{\nu} \approx mc^2 \Gamma/2$ is valid in any region. In
the opposite case, $\epsilon_{\nu} \approx (mc^2 \Gamma^2 \Gamma_{\rm n}^{-1} \sin^2
\theta) /4$ in regions (ii) and (iii).

The result of integration divided by the terminal proton energy $\Gamma_{\rm p} mc^2$
shows how large are the neutrino losses compared to the energy required to accelerate the
fireball:
\begin{equation} \label{nu-loss}
\delta E_{\nu} \approx 0.15\, \left( \frac{\Gamma_{\rm d}}{\Gamma_{\rm n}} \right)^3
\eta \tau_{\rm n} \times \left\{
\begin{array}{lr}
\displaystyle \frac{\Gamma_{\rm n}}{\Gamma_{\rm p}} \left( \ln \frac{\Gamma_{\rm
p}}{\Gamma_{\rm n}} +0.7 \right),& \sin \theta<\sqrt{2\Gamma_{\rm n}/\Gamma_{\rm
p}};\\ \displaystyle \sin^2 \theta + \frac{\Gamma_{\rm n}}{\Gamma_{\rm p}} \left( \ln
\frac{2}{\sin^2 \theta} -0.7 \right),& \sin\theta>\sqrt{2\Gamma_{\rm n}/\Gamma_{\rm
p}}.
\end{array} \right.
\end{equation}
Here $\tau_{\rm n} \approx 4$ is the optical depth of the decoupling region for neutrons
and the factor $(\Gamma_{\rm d}/\Gamma_{\rm n})^3$ is equal to about 2.5 rather than to a
value of $3^{3/2}$ corresponding to the decoupling point. This is explained by the fact
that the neutron flow continues to accelerate after the decoupling. Because of the
adopted approximations, Eq.~(\ref{nu-loss}) is only applicable if $\Gamma_{\rm
p}/\Gamma_{\rm n} \ga 4$. Note that if $\eta \approx 1$ and the magnetic field is
perpendicular to the radius, then over 50\% of the source power is lost due to neutrino
emission. This fact should be taken into account in calculations. The energy of a
neutrino for a typical source with $\Gamma_{\rm p} \sim 10^3$ and $\Gamma_{\rm
p}/\Gamma_{\rm n} \sim 4$ is about 30~GeV in the limiting case of a small angle $\theta$
and about $200\,\sin^2 \theta$~GeV in the opposite case.

\subsection{Spectrum of the electromagnetic cascade}

Comptonization of the thermal radiation near the photosphere proceeds in the classical
regime. Thus, if scattering of photons in the plasma is not taken into account, the flux
of the comptonized photons per unit energy has the power-law form
\begin{equation}
\label{em-spectr} F_{\varepsilon} \approx 0.2 \dot{N}_{\rm e} m_{\rm e} c^2 T^{-1/2}
\varepsilon^{-3/2},
\end{equation}
and cuts off at the energy $\varepsilon_{\max} \sim 4 \gamma_{\rm i}^2 T$. Here $\dot{N}_{\rm e}$ is the injection
rate of primary electrons and positrons,\footnote{The contribution of secondary $e^{-}e^{+}$ pairs becomes
important at energies less than $(\varepsilon_{\rm t}/ m_{\rm e}c^2)\,T$.} $\gamma_{\rm i}$ is their Lorentz
factor, and $m_{\rm e}$ is the electron mass. The scattering in thermal plasma results in the cutoff of the
initial spectrum at energies above $m_{\rm e}c^2/\tau$, where $\tau = \sigma_{\rm T} n_{\rm e} R/\Gamma$ is the
Thomson optical depth at a given radius and $n_{\rm e} \approx 0.4\,(\Gamma_{\rm p}/ \Gamma_{\rm n})\,[m_{\rm e}
c^2/(2T)]^{1/4}\,n_{\rm p}$ is the total number density of electrons and positrons determined by the balance
between the production and annihilation of $e^{-}e^{+}$ pairs. Strictly speaking, if $\tau < \varepsilon_{\rm
t}/(2 m_{\rm e} c^2)$, then higher-energy photons in the narrow transparency interval with the lower boundary near
$\varepsilon_{\rm t}/\tau$ can survive, but these photons are insignificant for the determination of the cascade
spectrum at energies below $ m_{\rm e} c^2$.

The radiation emitted at the maximum optical depth for a given photon energy dominates
the low-energy part of the electromagnetic-cascade spectrum ($\varepsilon \ll m_{\rm e}
c^2$). Such photons are subject to additional energy losses due to adiabatic expansion of
the plasma. In this case, the upper boundary of the spectrum varies in the same way as
the temperature of the background radiation and decreases by a factor of $(R_{\rm
T}/R)^{1/3}$, where $R_{\rm T}$ is the radius of the Thomson photosphere. Thus, the
spectrum at energies $\varepsilon \ll m_{\rm e} c^2$ is dominated by the radiation
emitted at $R\sim[\varepsilon/(m_{\rm e}c^2)]^{2/3}\,R_{\rm T}$. This radiation
corresponds to a photon flux per unit energy that is $(R_{\rm T}/R)^{1/3} = (m_{\rm e}
c^2/ \varepsilon)^{2/9}$ times greater than the contribution of photons emitted from the
Thomson photosphere. This yields the spectrum $F_{\varepsilon} \propto \varepsilon^q$
with the index $q=-31/18$.

The spectrum above $m_{\rm e} c^2$ observed at infinity is determined by the local radiation of the
electromagnetic cascade at the (energy-dependent) photosphere level. We point out three energy ranges. The opacity
at energies from $m_{\rm e} c^2$ to $\varepsilon_{\rm t}$ is determined by the Klein--Nishina scattering by
electrons and positrons, and the photospheric radius is equal to $R_{\rm ph} \approx (m_{\rm e} c^2/
\varepsilon)^{6/5}\,R_{\rm T}$. Since $\dot{N}_{\rm e} \propto R_{\rm ph}^{-1}$ in Eq.~(\ref{em-spectr}), we
arrive at the spectral index $q = -7/10$. If the photon energy exceeds $\varepsilon \approx \varepsilon_{\rm t}$,
the two-photon absorption becomes dominant, and the photosphere shifts away from the source as $R_{\rm ph} \approx
(m_{\rm e} c^2/\varepsilon_{\rm t}) (\varepsilon/\varepsilon_{\rm t})^{13/18}\,R_{\rm T}$ with increasing
$\varepsilon$. Here the exponent is determined by the spectral index in the region $\varepsilon < m_{\rm e} c^2$.
If $1 < \varepsilon/\varepsilon_{\rm t}<(\varepsilon_{\rm t}/ m_{\rm e} c^2)^{18/13} \approx 10$, the distance
from the source to the photosphere is less than $R_{\rm T}$, so that the radial variation in the temperature
should be taken into account, whereas $R_{\rm ph} > R_{\rm T}$ and $T= {\rm const}$ at higher photon energies.
Correspondingly, the spectral index $q=-107/54\approx -2$ in the first case and $q=-20/9$ in the second one.

The results obtained may be summarized as follows:
\begin{equation} \label{indices}
F_{\varepsilon} \propto \varepsilon^q\, , \qquad
\left\{
\begin{array}{ll}
q= -31/18, & \varepsilon < m_{\rm e} c^2;\\
q= -7/10, & m_{\rm e} c^2 < \varepsilon < \varepsilon_{\rm t}; \\
q= -107/54, & \varepsilon_{\rm t} < \varepsilon <
{\min} \{10\, \varepsilon_{\rm t},\varepsilon_{\max}\}; \\
q= -20/9, & {\min} \{10\, \varepsilon_{\rm t},\varepsilon_{\max}\}
< \varepsilon < \varepsilon_{\max}.\\
\end{array} \right.
\end{equation}
The main energy-carrying part of the spectrum (\ref{indices}) is the region between
$\varepsilon_{\rm t}$ and $\varepsilon_{\rm up}={\min} \{10\, \varepsilon_{\rm
t},\varepsilon_{\max}\}$. Multiplying by $\Gamma_{\rm p}$, we obtain an energy range of 2
to 50~GeV in the laboratory frame.

Applying the same method that was used to obtain Eq.~(\ref{nu-loss}), we calculate the fraction of energy $\delta
E_{\rm cas}$ corresponding to the emission of the electromagnetic cascade. In this case, the integration is
performed from $R_{\rm T}$ to infinity, the effective value of $\sin^2 \theta$ is equal to 1, and the obtained
result is multiplied by the ratio of the total energy of photons in the considered range to the energy of
electrons and positrons injected at the distance $R_{\rm T}$. This ratio is about $\ln (\varepsilon_{\rm
up}/\varepsilon_t)\,\sqrt{\varepsilon_{\rm up}/\varepsilon_{\rm max}}$. For large decoupling parameter we have
$\varepsilon_{\max}
> 10\, \varepsilon_{\rm t}$, which yields
\begin{equation} \label{cascade}
\delta E_{\rm cas} \approx 0.3\, \left( \frac{\Gamma_{\rm d}}{\Gamma_{\rm n}}
\right)^3 \left( \frac{R_{\rm s}}{R_{\rm T}} \right)
\sqrt{\frac{\varepsilon_t}{T}}\,\frac{\eta \tau_{\rm n}}{\gamma_{\rm i}}\, .
\end{equation}
The effective area of the gamma-ray telescope onboard the GLAST satellite will allow one to observe the
electromagnetic-cascade emission from gamma-ray bursts at redshifts $z \sim 0.1$ provided the burst energy is
$10^{52}$~erg.

The energy is partially emitted in the form of gamma-ray quanta directly coming from the
decay of neutral pions. These photons fall in the 100-GeV energy range in the laboratory
frame. The fraction of energy carried by such a radiation is given by the following
formula obtained in~\cite{apj}:
\begin{equation} \label{E-gamma}
\delta E_{\gamma} \approx \frac{2\varepsilon_{\rm i} \sigma_{\pi 0} \Gamma_{\rm p}
\eta n_{\rm p} \varepsilon_{\rm t}} {m \sigma_{\rm T} \Gamma_{\rm n} n_{\rm e} m_{\rm
e}} \left( \frac{\varepsilon_{\rm t}}{\varepsilon_{\rm i}} \right)^{0.72}\sim(1\!-\!2)
\cdot 10^{-3} \eta \left( \frac{\Gamma_{\rm p}}{\Gamma_*} \right)^{7/30}.
\end{equation}
Here $\varepsilon_{\rm i}$ is the average energy (in the fireball frame) of gamma-quanta born in the decay of
neutral pions. This formula is corrected for the more detailed model of the electromagnetic-cascade spectrum
developed above. However, we find that the numerical value of $\delta E_{\gamma}$ is almost unaffected by the
specific form of the cascade spectrum. This value is changed by a factor less than 1.3 compared to the simplified
model. The photons from $\pi_0$ decay can be detected by ground-based \v{C}erenkov telescopes in the near future
when such facilities as MAGIC or ARGO become fully operational.

\section{Conclusions}

It is shown above that, in certain cases, free neutrons are inevitably created in
astrophysical sources of X- and gamma-ray radiation due mainly to dissociation of helium
nuclei. A temperature of 0.7~MeV required for this process is either achieved in the
source initially (gamma-ray bursts, inner regions of accretion disks in
advection-dominated models) or reached due to the large relative velocity of seed
neutrons and ions (e.g., jets in microquasars and SGRs).

A neutron halo in the inner region of an accretion disk originates owing to the effect of
neutron pile-up. In this case, the radial velocity of the neutron component is usually
much less compared to the electron-ion plasma, while the temperature of this component is
high enough to ensure collisional dissociation of helium nuclei. Neutron-neutron
collisions give rise to the neutron wind whose dynamical pressure is enough to terminate
accretion from the wind of the companion star. As the accretion is terminated, the
neutron wind weakens, the accretion starts again, and the entire automodulation cycle of
the accretion rate is repeated. A fraction of neutrons from the wind is intercepted by
the cold outer disk where these neutrons lead to the formation of nuclear lines,
primarily the 2.2-MeV deuterium line. In the immediate vicinity of the central object,
the neutrons undergo inelastic collisions and produce pions, thus initiating the
electromagnetic cascade which finally produces the comptonized radiation with a hard
power-law spectrum and the annihilation line at 511~keV generated by positron
annihilation.

In the course of acceleration of a plasma flow comprising a neutron component, the
neutrons accelerate due only to their collisions with protons. As soon as the time
between collisions exceeds the characteristic time scale of the fireball expansion, the
acceleration of neutrons terminates. Thus, the Lorentz factors of the neutron and proton
components can become strongly different. Such a decoupling results in the dissociation
of helium nuclei (if helium exists in the flow) due to interaction with seed neutrons and
leads to an exponential increase in the number of neutrons. Moreover, the decoupling
gives rise to inelastic nucleon collisions in which about one pion per neutron is born.
About a half of the pion energy is lost for neutrino emission, which has an important
effect on the energy balance of the source. Pion decay gives rise to an electromagnetic
cascade in which a significant amount of electron-positron plasma is created. Due to the
cascade, the fireball photosphere shifts outward, and hard gamma-ray emission appears.
Observations of the spectral features of the electromagnetic cascade allow one to
determine the plasma parameters and can be indirect evidence for the existence of
neutrons in the fireball.

The above-mentioned effects are also possible in the absence of decoupling if the
fireball is strongly inhomogeneous (variability time of the source about $R_{\rm g}/c$)
or strongly anisotropic (with a scale about $1/\Gamma$) and its Lorentz factor exceeds
$\Gamma_*^{4/5}$. In this case, the neutrino losses decrease, while the
electromagnetic-cascade spectrum remains almost unchanged.

Finally, since neutrons carry a significant fraction of the fireball energy which is
converted to radiation only upon neutron decay, they can strongly affect the light curves
of transient sources, including the formation of a secondary maximum and a bimodal
distribution over the duration~[14, 15].

\Acknowledgements

This work is supported by the Russian Foundation for Basic Research (project
\#~99--02--18244) and the Commission on Work with Youth of the Russian
Academy of Sciences (project ``Interaction of powerful electromagnetic and
neutrino radiation with plasmas in superstrong magnetic and gravitational
fields of compact astrophysical objects: white dwarfs, neutron stars, and
black holes'').


\begin{thebibliography}{99}


\bibitem{CK}
J.~Contopoulos and D.~Kazanas, {\em Astrophys.\ J}., {\bf 441}, 521 (1995).
\bibitem{s76}
S.\,L.~Shapiro, A.\,P.~Lightman, and D.\,M.~Eardley, {\em  Astrophys.\ J.},  {\bf 204},
187 (1976).
\bibitem{N96}
R.~Narayan, J.~McClintock, and I.~Yi, {\em  Astrophys.\ J.}, {\bf 457}, 821 (1996).
\bibitem{AS}
F.\,A.~Aharonyan and R.\,A.~Sunyaev, {\em  MNRAS}, {\bf 210}, 257 (1984).
\bibitem{G89}
N.~Guessoum and D.~Kazanas, {\em  Astrophys.\ J.}, {\bf 345}, 356 (1989.
\bibitem{6}
N.~Guessoum and D.~Kazanas, {\em  Astrophys.\ J.}, {\bf 358}, 525 (1990).
\bibitem{MR}
I.\,F.~Mirabel and L.\,F.~Rodr\'{i}guez, {\em  Nature}, {\bf 392}, 673 (1998).
\bibitem{rev1}
K.\,A.\,Postnov, {\em Uspekhi Fiz.\ Nauk}, {\bf 169}, 545 (1999).
\bibitem{rev2}
T.~Piran {\em  Phys.\ Rep.}, {\bf 314}, 575 (1999).
\bibitem{Opt1}
M.\,R.~Metzger, S.\,G.~Djorgovski, S.\,R.~Kulkarni, et.al., {\em  Nature}, {\bf 387} 878
(1997).
\bibitem{Opt2}
J.\,P.~Halpern, J.\,R.~Thorstensen, D.\,J.~Helfand, and E.~Costa, {\em  Nature}, {\bf
393}, 41 (1998).
\bibitem{Opt3}
S.\,R.~Kulkarni, S.\,G.~Djorgovski, S.\,C.~Odewahn, et.al., {\em  Nature}, {\bf 398} 389
(1999).
\bibitem{gamma}
M.\,G.~Baring and A.\,K.~Harding, {\em  Adv. Space Res.}, {\bf 15}, No~5, 153 (1995).
\bibitem{aa}
E.\,V.~Derishev, V.\,V.~Kocharovsky, and Vl.\,V.~Kocharovsky, {\em  Astron.\ Astrophys.},
{\bf 345}, L51 (1999).
\bibitem{apj}
E.\,V.~Derishev, V.\,V.~Kocharovsky, and Vl.\,V.~Kocharovsky, {\em  Astrophys.\ J.}, {\bf
521}, 640 (1999).
\bibitem{Hyp1}
S.\,E.~Woosley, {\em  Astrophys.\ J.}, {\bf 405}, 273 (1993).
\bibitem{Hyp2}
B.~Paczy\'{n}ski, {\em  Astrophys.\ J.}, {\bf 494}, L45 (1998).
\bibitem{indColl}
E.\,V.~Derishev, V.\,V.~Kocharovsky, and Vl.\,V.~Kocharovsky, {\em Radiophys.\ Quantum Electron}.,
 {\bf 41}, 7 (1998).
\bibitem{18}
E.\,V.~Derishev, V.\,V.~Kocharovsky, and Vl.\,V.~Kocharovsky, {\em JETP Lett}., {\bf 70}, 652 (1999).
\bibitem{NsNs}
B.~Paczy\'{n}ski {\em  Astrophys.\ J.}, {\bf 363}, 218 (1990).
\bibitem{ShT}
S.\,L.\,Shapiro and S.\,A.Teukolsky, {\em Black Holes, White Dwarfs, and Neutron Stars}, John Wiley and Sons, New
York (1985).

\end{thebibliography}
\end{document}